\documentclass[conference,letterpaper]{IEEEtran}
\usepackage{amsmath} 
\usepackage{hyperref}
\usepackage{amssymb}
\usepackage{graphicx} 
\usepackage{tikz}  
\usetikzlibrary{positioning, shapes.geometric} 
\usepackage{pgfplots}
\pgfplotsset{compat=1.18} 
\usepackage{csvsimple}
\usepackage{makecell}
\usepackage{algorithm2e} 
\usepackage{array}
\usepackage{multirow}
\usepackage{longtable}
\usepackage{rotating}

\newcommand{\Alphabet}[1]{\Sigma_{#1}}

\newcommand{\subOp}{\textbf{SUB}}
\newcommand{\insOp}{\textbf{INS}}
\newcommand{\delOp}{\textbf{DEL}}

\newcommand{\subOpF}[2]{\subOp({#1},{#2})}
\newcommand{\insOpF}[1]{\textbf{INS}({#1})}
\newcommand{\delOpF}[1]{\textbf{DEL}({#1})}

\newcommand{\ECG}[2]{\textbf{ECG}({#1},{#2})}
\newcommand{\LossFunc}[1]{\textbf{Loss}({#1})}

\newtheorem{definition}{Definition}[section]
\newtheorem{example}{Example}[section]
\newtheorem{proposition}{Proposition}[section]
 
\title{Unrestricted Error-Type Codebook Generation for Error Correction Code in DNA Storage Inspired by NLP} 

\author{
  \IEEEauthorblockN{Yi Lu\IEEEauthorrefmark{1},
                    Yun Ma\IEEEauthorrefmark{1},
                    Chenghao Li\IEEEauthorrefmark{2},
                    Xin Zhang \IEEEauthorrefmark{3},
                    Guangxiang Si\IEEEauthorrefmark{4}}
  \IEEEauthorblockA{\IEEEauthorrefmark{1}
                   School of Mathematical Sciences, Capital Normal University, Beijing, China,
                   Email:\{2230501004,2210502084\}@cnu.edu.cn}
  \IEEEauthorblockA{\IEEEauthorrefmark{2}
                    Liangjiang International College, Chongqing University of Technology, Chongqing, China,
                    Email:lch17692405449@gmail.com }
  \IEEEauthorblockA{\IEEEauthorrefmark{3}
                    Institute of Systems Biomedicine, Peking University, Beijing, China
                    Email: 18830053918@163.com}
  \IEEEauthorblockA{\IEEEauthorrefmark{4}
                   School of Biological Science and Medical Engineering, Southeast University, Nanjing, China.
                   Email:guangxiang\_si@163.com  }
}

\begin{document} 
\maketitle

\begin{abstract}
THIS PAPER IS ELIGIBLE FOR THE STUDENT PAPER AWARD.\\
 Recently, DNA storage has surfaced as a promising alternative for data storage, presenting notable benefits in terms of storage capacity, cost-effectiveness in maintenance, and the capability for parallel replication. Mathematically, the DNA storage process can be conceptualized as an insertion, deletion, and substitution (IDS) channel. Due to the mathematical complexity associated with the Levenshtein distance, creating a code that corrects for IDS remains a challenging task. In this paper, we propose a bottom-up generation approach to grow the required codebook based on the computation of Edit Computational Graph (ECG) which differs from the algebraic constructions by incorporating the Derivative-Free Optimization (DFO) method. Specifically, this approach is regardless of the type of errors. Compared the results with the work for 1-substitution-1-deletion and 2-deletion, the redundancy is reduced by about 30-bit and 60-bit, respectively. As far as we know, our method is the first IDS-correcting code designed using classical Natural Language Process (NLP) techniques, marking a turning point in the field of error correction code research. Based on the codebook generated by our method, there may be significant breakthroughs in the complexity of encoding and decoding algorithms.
\end{abstract} 

\section{Introduction}
\subsection{Background and Motivation}

DNA, as the most natural form of information storage, presents numerous advantages over traditional storage mediums\cite{Dong_Sun_Ping_Ouyang_Qian_2020,organick2018random}, notably its high storage density and low maintenance costs, which establish it as an ideal medium for long-term information storage. Due to the characteristics of the DNA storage medium, it is susceptible to environmental factors and experimental errors, leading to data inaccuracies. Therefore, error correction codes become crucial in DNA storage\cite{Smagloy_Welter_Wachter-Zeh_Yaakobi_2020,Cai_Chee_Gabrys_Kiah_Nguyen_2021}.

Despite advancements in correcting deletion errors, challenges persist \cite{Sima_Raviv_Schwartz_Bruck_2023}. These include constructing minimal-redundancy deletion codes that allow efficient decoding, improving existential redundancy bounds, and efficiently addressing a mix of deletions and substitutions. Although some research \cite{Song_Cai_2022,Smagloy_Welter_Wachter-Zeh_Yaakobi_2020} has proposed codes combining the mentioned deletion and substitution codes, the quest for more redundancy-efficient methods continues.

\vspace{12pt}

The string-to-string correction problem \cite{Wagner_Fischer_1974,Damerau,Kukich_1992} as a tpoic in NLP has been extensively researched in the past and plays a crucial role in various fields, just as we have seen, such as, spell checking, auto-correction, search engines and information retrieval. In NLP tasks, at least a correct corpus is needed as an evaluation criterion, but there is no such criterion in coding theory. We only need to consider how to efficiently generate the codebook.

Currently, there is little literature \cite{Wu_2018, Guo_Sun_Wei_Wei_Chen_2023} discussing the generation of codebooks. However, the method in literature \cite{Guo_Sun_Wei_Wei_Chen_2023} is only designed to address errors related to edit distance. Moreover, it is challenging to generate long sequences through top-down filtering. 
While the approach in literature \cite{Wu_2018} focuses exclusively on substitution-type errors. The codebook generated by our method is capable of handling errors involving p-substitution-q-deletion. Furthermore, this method has the potential to generate a codebook tailored to the maximum errors for different nucleotides.

\subsection{An Overview of Our Solution}

A bottom-up approach to generate the codebook by iteration is adopted, avoiding inefficient brute-force searches in $4^n$ arbitrary sequences. Utilizing derivative-free optimization (DFO) methods \cite{Larson_Menickelly_Wild_2019}, we optimize redundancy to an acceptable range while meeting error correction requirements. Once the codebook is obtained, there is no need to generate a new one unless there is a demand for improvement in the current codebook. We can encode the input binary sequence segment-wise based on the codebook index. The decoding and error correction processes have been extensively studied in classical literature \cite{Kashyap_Oommen_1981,Peterson_1980,Oommen_Loke_1997,Oommen_Loke_1995}. Therefore, this paper will focuse solely on the generation of the codebook.

In order to generate the desired codebook, the first step is to design a "loss function" called ECG, that reflects how much the current codebook achieve the need of error correction in each step. For the overall loss, we can consider it as the sum of the losses between each pair of distinct sequences. When designing this algorithm, we need to ensure that it satisfies the following points:
\begin{itemize}
    \item \textbf{Residues of Feasible Edit Counts}. Differing from sequence matching and sequence similarity, what we consider here is all feasible edits between two sequences, providing sufficient indication for the redundancy to be added in each iterative step. The entire computation is still implemented based on dynamic programming \cite{Wagner_Fischer_1974}. At the same time, it is necessary to return a non-negative value for residues, which is used to determine how many operations are remaining until the completion of the distance calculation.
    \item \textbf{Constrained Edit Information}. This idea \cite{Oommen_1986} takes into account the impact of the maximal number of substitution, insertion, and deletion on the computation of edit information. Due to the varied distribution of errors associated with different nucleotides \cite{organick2018random}, a more stringent constraint, considering the limitations on different types of editing counts for various nucleotides, is worth considering. This implies a reduction in the volume of the error ball, thereby holding great potential for enhancing the redundancy of the encoding.
    \item \textbf{Incremental Computation}. Increasing efficiency as the string length grows, we have the concept of incremental computation for edit distance \cite{Kim_Park_2000}. This involves calculating additional edit information based solely on the previous step's results, rather than re-evaluating from the start. 
    \item \textbf{Lazy Evaluation}. Although insertions and deletions can affect the positions of sequence matches, once the upper bounds for insertion and deletion counts are determined, it is unnecessary to calculate all positions \cite{ALLISON1992207}. 
\end{itemize}

Based on the ECG mentioned above, after defining the maximum error scenario and codebook size, we can iteratively generate the codebook. We start by initializing a set of empty sequences. Next, for each sequence, we enter a loop of two steps: \textbf{generation of candidate}, \textbf{selection of the optimal candidates} and  result concatenation,
\begin{itemize}
    \item Generation of candidate: In DNA storage, there is often a requirement for \textbf{compatibility}, such as GC balance and run length. This requirement can be addressed during candidate generation \cite{Knuth_1986,Liu_He_Tang_2022}.
    \item Selection of the optimal candidates: Due to the discrete nature of sequences, the process of selecting the optimal candidates for maximal code rate should involve DFO .
\end{itemize}
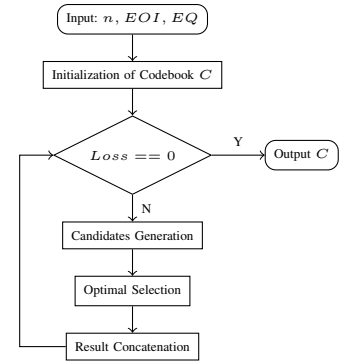
\begin{figure}[htbp]
    \centering
    \begin{tikzpicture} [scale=0.5,node distance=10pt,auto]
         \node[draw, rounded corners]                        (start) [font=\tiny]  {Input: $n,EOI,EQ$};
  \node[draw, below=of start]                         (step1) [font=\tiny] {Initialization of Codebook $C$};
  \node[draw, diamond, aspect=2, below=of step1]     (choice) [font=\tiny] {$Loss==0$};
  \node[draw, below= of choice]                   (step2)[font=\tiny]  {Candidates Generation};
  \node[draw, below= of step2]                   (step3) [font=\tiny] {Optimal Selection};
  \node[draw, below= of step3]                   (step4) [font=\tiny] {Result Concatenation};
  \node[draw, rounded corners, right=20pt of choice]  (end)   [font=\tiny]  {Output $C$};
  
  \draw[->] (start)  -- (step1);
  \draw[->] (step1) -- (choice);
  \draw[->] (choice) --node[above,font=\tiny]{Y} (end);
  \draw[->] (choice) --node[right,font=\tiny]{N} (step2);
  \draw[->] (step2) -- (step3);
  \draw[->] (step3) -- (step4);
  \draw[->] (step4) -- ++(-3,0) |- (choice); 
    \end{tikzpicture}
    \caption{Workflow of codebook generation}
    \label{fig:workflowOfCodeGeneration}
\end{figure}

\subsection{Arrangement}
We outline the section arrangement of this paper.

In the next section\ref{sec:basicknowledge}, we introduce the concepts and symbols relevant to this paper. In Section\ref{sec:ECGandBitEncoding}, the ECG (Edit Computational Graph) algorithm and its implementation using bitwise operations are introduced. Then, we present the process of encoding generation. In this paper, the focus is primarily on utilizing the Monte Carlo method as the optimization algorithm. In Section\ref{sec:analysis}, we provide a rough analysis of its time and space complexity. We conclude the paper in Section\ref{sec:Conclu}. 
\section{Basic Concepts and Notations}
\label{sec:basicknowledge}

In this paper, we state several symbols: 
\begin{center}
    \begin{tabular}{p{3cm}<{\centering}|p{5cm}<{\centering}}
     Symbol & Meaning \\ 
     \hline
     $\Alphabet{q}$  & q-alphabet with q elements.\\
     C & A codebook over $\Alphabet{q}$ .\\
     $l(C)$ & The length of each sequence in $C$.\\ 
     $\lvert C\rvert$ & Size of codebook.\\
     $n(C)$ & The number of binary bits which $C$ can represent .\\
     $r(C)$ & The redundancy of C.
\end{tabular}
\end{center}

Concretely, we use the symbols $\Sigma_2=\{0,1\}$ and $\Sigma_{4}=\{A,G,C,T\}$ to represent alphabets of binary symbols and nucleotide symbols, respectively.

Naturally, a metric is a measure of the degree of distinction between two sequences.
\begin{definition}
    The \textbf{metric} of a sequence is the following mapping
    \[d:\Sigma^n\times \Sigma^n\rightarrow \mathbb{R}_{\ge0},\]
    satisfying the following property: $\forall x,y,z\in\Sigma^n$
    \begin{itemize}
        \item \textbf{Non-negativity}: $d(x,y)\ge0$.
        \item \textbf{Identity}: $d(x,y)=0\Leftrightarrow x=y.$
        \item \textbf{Symmetry}: $ d(x,y)=d(y,x).$
        \item \textbf{Triangle Inequality}: $d(x,y)\le d(x,z)+d(z,y).$
    \end{itemize}
\end{definition}

Specifically, there are several types of metric, 

We can define a "ball" in sequence metric space.
\begin{definition}
    A(n) (error) \textbf{ball} in the code space of sequence $s$ and radius $r$ is defined as follow:
    \[B(s;r)=\{s'\in \Sigma^{|s|}:d(s,s')\le r\}.\]
\end{definition}
For any $C$, if it exhibits a low enough level of ambiguity, it implies a certain error-correcting capability. Thus for a given number $r$ and metric $d$, the following conditions will be satisfied:
\begin{equation}
    B(s_1;r)\cap B(s_2;r)=\emptyset,\forall s_1,s_2\in C,s_1\ne s_2.
    \label{eq:intersectEmptyCond}
\end{equation}
Therefore, we have the following proposition.
\begin{proposition}
    Let $C$ be a codebook. $r$ and $d$ are the same as above and satisfy the Condition \ref{eq:intersectEmptyCond}. Then,
    \begin{itemize}
        \item If $d=d_H$,  $C$ possesses the capability to correct $r$ substitution errors, i.e. $$\forall s_1,s_2\in C,d_H(s_1,s_2)\ge 2r+1.$$
        \item If $d=d_L$,  $C$ possesses the capability to correct $r$ edit (the sum of substitution, insertion, and deletion errors) errors, i.e. $$\forall s_1,s_2\in C,d_L(s_1,s_2)\ge 2r+1.$$
        \item If $d=d_{LCS}$, $C$ possesses the capability to correct $r$ insertion, and deletion (sum) errors, i.e. $$\forall s_1,s_2\in C,d_{LCS}(s_1,s_2)\ge 2r+1.$$
    \end{itemize}
\end{proposition}

However, the concept of a metric replaces dissimilarity with numerical values, much like a “projection”, leading to the loss of a substantial amount of information.
\begin{example}
     We define a codebook $C\subset \Sigma_4^{10}$ to be capable of correcting simultaneous occurrences of $2$ substitution errors, $1$ insertion error and $1$ deletion error. In fact, for $s_1=TCTTCTTCCG,s_2=TCCGCAGAAT$, $s_1,s_2\in C$, we cannot edit $s_1$ to $s_2$ within $4sub-2ins-2del$. However, $d_L(s_1,s_2)= 7$, and if we take the Levenshtein distance, \[B(s_1;4)\cap B(s_2;4)\ne\emptyset.\]
\end{example}

Therefore, we cannot simply combine all editing methods into a single value for analysis.

Let's first introduce several symbols to represent different editing methods.
\begin{definition}
    All of the operations are based on $\Alphabet{4}$.We use $[\subOp],[\insOp],[\delOp]$ to present the class of substitution, insertion and deletion respectively. In each class, we have
    \begin{itemize}
        \item $[\subOp]=\{\subOpF{c_1}{c_2}:c_1,c_2\in \Alphabet{4}\}\cup \{\subOpF{*}{*}\},$
        \item $[\insOp]=\{\insOpF{c}:c\in\Alphabet{4}\}\cup\{\insOpF{*}\},$
        \item $[\delOp]=\{\delOpF{c}:c\in\Alphabet{4}\}\cup\{\delOpF{*}\},$
    \end{itemize}
    where $*$ means the absence of distinction of specific characters. 
\end{definition}


\begin{definition}
    \textbf{Edit of Interest (EOI)} is defined as an ordered array comprising elements in $[\subOp]$, $[\insOp]$, and $[\delOp]$ classes, arranged in sequential order.

    Correspondingly, the maximum occurrence count for each edit is referred to as the \textbf{Edit Quota (EQ)}. $EQ$ is represented as an array, where each position aligns with the respective operation in $EOI$.

    \textbf{Feasible Edit Count (FEC)} refers to the tuple representing the count of feasible edits generated during the editing process for a given $EOI$ and $EQ$.
\end{definition}

\begin{example}
    Let $s_1=AGC,s_2=AGG$, and $EOI=([\subOp],[\insOp],[\delOp]),EQ=(1,1,1)$. So the feasible edit count from $s_1$ to $s_2$ is $(1,0,0),(0,1,1)$.
\end{example}

Generally, we have the following proposition:
\begin{proposition}
    Given $EOI,EQ$, for a codebook $C$ to correct errors defined by $EQ$, any two sequences, $s_1,s_2\in C,s_1\ne s_2$, must satisfy: it is impossible to edit $s_1$ into $s_2$ within $2*EQ$ operations or fewer.
\end{proposition}
This proposition is the generalization of Condition \ref{eq:intersectEmptyCond}.
The question of calculating the "edit reachability" within $EQ$ and effective calculation methods will be discussed in the next section.

\section{Edit Computational Graph and Bit Encoding}
\label{sec:ECGandBitEncoding}
The name "Edit Computational Graph" comprises two elements. Firstly, "edit" emphasizes the specific editing operations under consideration. Secondly, "computational graph" refers to a Directed Acyclic Graph (DAG), a term commonly used in machine learning. "Graph" essentially offers a statement of the workflow for our algorithm. In this context, it signifies the use of a similar DAG structure for computing information related to edits.

\subsection{Introduction of ECG}
\begin{definition}
    Given $n\in\mathbb{Z}_{>0}$, $EOI$ (Edit of Interest) and $EQ$ (Edit Quota), along with any two sequences $s_1,s_2\in\Alphabet{4}^n$, we denote their \textbf{Edit Computational Graph} as $\ECG{s_1}{s_2;EOI,EQ}$ or $\ECG{s_1}{s_2}$ if there is no ambiguity. The ECG consists of two components: vertices ($V$) and edges ($E$).

    Each node can be labeled with $(i, j)$, indicating all FECs within $EQ$ used to transform $s_1[:i]$ to $s_2[:j]$.

    Each edge represents different transitions for all FECs.    
    For node $(i, j)$, let $[\epsilon]$ be nothing to do, we have
    \[\begin{cases}
        (i-1,j-1)\rightarrow (i,j):\epsilon,& s_1[i]==s_2[j];\\
        \begin{cases}
            (i-1,j-1)\rightarrow (i,j):[\subOp],\\
            (i-1,j)\rightarrow (i,j):[\delOp],\\
            (i,j-1)\rightarrow (i,j):[\insOp],
        \end{cases}&s_1[i]!=s_2[j].
    \end{cases}\]
    As illustrated in the Figure\ref{fig:StateTrans}. 
  
    \begin{figure}[htbp]
        \centering
        \begin{tikzpicture} [scale=0.5]
            \fill[black] (2,-0.5) circle (2pt) node[below,font=\footnotesize]{$(i-1,j)$};
            \fill[black] (0,0) circle (2pt) node[below,font=\footnotesize]{$(i,j)$};
            \fill[black] (-2,-0.5) circle (2pt) node[below,font=\footnotesize]{$(i,j-1)$};
            \fill[black] (0,2) circle (2pt) node[above,font=\footnotesize]{$(i-1,j-1)$};
            \draw[->, >=stealth, thick ] (2,-0.5) --node[above right,font=\footnotesize]{Deletion} (0.1,0) ;
            \draw[->, >=stealth, thick] (-2,-0.5) --node[above left,font=\footnotesize]{Insertion} (-0.1,0) ;
            \draw[->, >=stealth, thick] (0,2) --node[left,font=\footnotesize]{Substitution} (0,0.1) ;
            \node at (0,-2) [font=\footnotesize]{(a) When $s_1[i] != s_2[j]$};
        \end{tikzpicture}    
        \begin{tikzpicture} [scale=0.5]
            \fill[black] (2,-0.5) circle (2pt) node[below,font=\footnotesize]{$(i-1,j)$};
            \fill[black] (0,0) circle (2pt) node[below,font=\footnotesize]{$(i,j)$};
            \fill[black] (-2,-0.5) circle (2pt) node[below,font=\footnotesize]{$(i,j-1)$};
            \fill[black] (0,2) circle (2pt) node[above,font=\footnotesize]{$(i-1,j-1)$}; 
            \draw[->, >=stealth, thick] (0,2) --node[left,font=\footnotesize]{No error} (0,0.1) ;
            \node at (0,-2) [font=\footnotesize]{(b) When $s_1[i] == s_2[j]$};

        \end{tikzpicture} 
        \caption{State transitions on different edges. (a) represents the case of $s_1[i]!=s_2[j]$ and (b) shows the case of  $s_1[i]==s_2[j]$.}
        \label{fig:StateTrans}
    \end{figure}
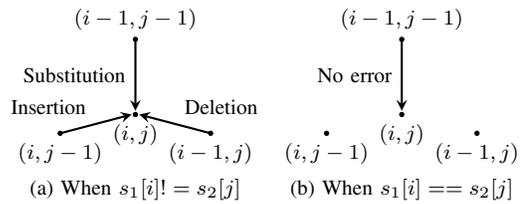
\end{definition}

\begin{example}
    \label{eg:DemoTransByEdge}
    We assume $EOI=\{\subOp, \insOp, \delOp\}$ and $EQ=(4,2,2)$. Suppose $s_1[i] !=s_2[j]$ and the data at each node are:
    \begin{itemize}
        \item $(i-1,j):\{(0,0,0),(4,1,1),(1,2,1),(1,1,2)\};$
        \item $(i,j-1):\{(0,0,0),(4,1,1),(1,2,1),(1,1,2)\};$
        \item $(i-1,j-1):\{(0,0,0),(4,1,1),(1,2,1),(1,1,2)\};$
    \end{itemize}
    From $(i-1,j)$ to $(i,j)$, add $1$ to the coordinates corresponding to $\delOp$ and add it to $(i,j)$. Then
    \[(i,j):\{(0,0,1),(4,1,2),(1,2,2)\}.\]
    Since $(1,1,3)$ is overflow, it is not added. 
    
    Next, from $(i,j-1)$ to $(i,j)$, we have
    \[(i,j):\{(0,0,1),(4,1,2),(1,2,2),(0,1,0),(4,2,1)\}.\]
    Note that the set does not retain duplicate elements.

    Finally, from $(i-1,j-1)$ to $(i,j)$, we have
    
    \[\begin{aligned}
        (i,j):\{&(0,0,1),(4,1,2),(1,2,2),\\
                &(0,1,0),(4,2,1),(1,0,0),\\
                &(2,2,1),(2,1,2)\}.
    \end{aligned}\]
    Furthermore, if $s_1[i]==s_2[j]$, then
    $$(i,j):\{(0,0,0),(4,1,1),(1,2,1),(1,1,2)\},$$
    which is same as $(i-1,j-1)$.
\end{example}

The case of such nodes may not happen in reality, but this is just an intuitive example to demonstrate the data flow. Through the above example, we summarize the following key points that the ECG should have.
\begin{itemize}
    \item \textbf{Finiteness}: The number of all FECs retained by each node is finite, at most not exceeding
    $L=\prod_i(EQ[i]+1).$
    \item \textbf{Non-redundancy}: Each node retains non-repeating FECs, meaning that all FECs held by each node are distinct.
    \item \textbf{Boundedness}: For any FEC $\alpha$ and any $i$ in $[|EQ|]$, $\alpha[i]$ is nonnegative and bounded by $EQ[i].$ Otherwise, discard FECs with overflowed coordinates.
    \item \textbf{Edge-Dependency}: For any FEC, the updated coordinates are only related to the traversed edges (Figure\ref{fig:StateTrans}).
    \item \textbf{Monotonicity}: For any FEC, whenever passing through an edge with operations ($s_1[i]!=s_2[j]$), there must be an increase in certain coordinates.
\end{itemize}

\subsection{Bit Encoding and Algorithms}
In order to efficiently achieve the aforementioned objectives, we encode all of the FEC of a node using a bit with length $L$, and corresponding state transitions are implemented using bitwise operations.

\begin{definition}
    \label{def:IndEnDe}
    Given base $EQ$, we define \textbf{index encode} \[\varphi:[L]\rightarrow \prod_{i\in [|EQ|]}[EQ[i]]\] and \textbf{index decode} \[\psi:\prod_{i\in [|EQ|]}[EQ[i]]\rightarrow [L]\] as a pair of mutually inverse bijections, where
    \[\varphi(i)[k]=\begin{cases}
        i \mod{EQ[0]},& k=0\\
        \lfloor i/(\prod_{j=0}^{k-1}EQ[j])\rfloor\mod{EQ[k]},& \text{otherwise}
    \end{cases},\]
    and assume $(a_0,a_1,...,a_{|EQ|})\in \prod_{i\in [|EQ|]}[EQ[i]]$
    \[\psi((a_0,a_1,...,a_{|EQ|}))=a_0+\sum_{i=1}^{|EQ|-1}a_i(\prod_{j=0}^{i-1} EQ[j]).\]
\end{definition}
For the node $(i, j)$ , all its FECs are denoted as $\beta(i, j)\in\Alphabet{2}^L$, satisfying the following conditions:
\[\beta(i,j)[k]=\begin{cases}
    1 & \varphi(i)\in (i,j)\\
    0 & \varphi(i)\notin (i,j)
\end{cases}.\]

\begin{example}
    Recall the Example\ref{eg:DemoTransByEdge}. $L=(4+1)(2+1)(2+1)=45$ and $\beta(i,j)=01010000\;01000000\;01000001\;01000000\;00000000\;01010\in \Alphabet{2}^L,$
    when $s_1[i]!=s_2[j]$.
\end{example}

\begin{definition}
    \label{def:EMandSA}
    The \textbf{Edit Mask (EM)} and the \textbf{Shift Amount (SA)} associated to $EOI,EQ$ are two list defined below:
    \[EM[i]\in \Alphabet{2}^L, EM[i][k]=\begin{cases}
        1&\varphi(k)[i] < EQ[i],\\
        0&\varphi(k)[i] == EQ[i],
    \end{cases}\]
    \[SA[i]=\begin{cases}
        1&i=0,\\
        SA[i-1]*(EQ[i-1]+1)&i>0.
    \end{cases}\]
\end{definition}

Based on Definition\ref{def:IndEnDe} and Definition\ref{def:EMandSA}, we can provide a description of the updated state transition algorithm for FEC (Algorithm\ref{alg:TransOfFEC}):

For the bit array of each node $\beta\in\Alphabet{2}^L$, we introduce a loss function here to facilitate the latter generation algorithm.
\begin{definition}
    Given $EOI,EQ$, a loss function of bit array with regard to all FECs of a node $\LossFunc{\beta}$ is defined as follows:
    \[\begin{aligned}
        \textbf{Loss}:\Alphabet{2}^L&\rightarrow [L]\cup\{L\}\\
        \beta&\mapsto L-min\{i:\beta[i]==1\}.
    \end{aligned}\]
\end{definition}
\begin{example}
    Fix $s_1,s_2,EOI,EQ$, 
    \begin{itemize}
        \item if $s_1==s_2$, then $\beta(n,n)[0]=1,\LossFunc{\beta(n,n)}=L;$
        \item if $s_1$ cannot be edited into $s_2$ under the $EQ$ condition, then $\forall k\in [L],\beta(n,n[k]=0,\LossFunc{\beta(n,n)}=0.$
    \end{itemize}
\end{example}

In coding theory, insertion and deletion edits significantly alter sequence structure, so it is necessary to compute $q=max\{|[\insOp]|,|[\delOp]|\}$. 

\begin{figure}[htbp]
    \centering 
        \includegraphics[scale=0.5]{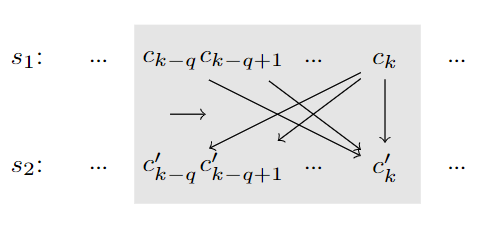} 
    \caption{Sequence alignment procedure in ECG at step $k$. If $k<q$, we just start from the index $0$.}
    \label{fig:DemoSeqAlign}
\end{figure}

As a consequence, when calculating the ECG, it is also not necessary to compare every position of the two sequences, as illustrated in Figure\ref{fig:DemoSeqAlign}.

As a variant of sequence alignment algorithm, its basic idea is based on the dynamic programming algorithm. For the dynamic programming matrix, we implement a more space-efficient approach as illustrated in the figure.
\begin{figure}
    \centering
    \begin{tikzpicture}[scale=0.6]
        \draw[dashed] (-1.2,-0.4) rectangle (1.2,0.4);
        \draw[dashed] (-4,-0.4) rectangle (-1.2,0.4);
        \draw[dashed] (-6,-0.4) rectangle (-4,0.4);
        \draw[dashed] (1.2,-0.4) rectangle (4,0.4);
        \draw[dashed] (4,-0.4) rectangle (6,0.4);
        
        \draw[dashed] (-1.2,-1.2) rectangle (1.2,-0.4);
        \draw[dashed] (-4,-1.2) rectangle (-1.2,-0.4);
        \draw[dashed] (1.2,-1.2) rectangle (4,-0.4);
        \draw[dashed] (-6,-1.2) rectangle (-4,-0.4);
        \draw[dashed] (4,-1.2) rectangle (6,-0.4);

        \node (a0) at (0,0)[font=\tiny] {$(k,k)$};
        \node (a1) at (2.5,0)[font=\tiny] {$(k-1,k)$};
        \node (a2) at (-2.5,0)[font=\tiny] {$(k,k-1)$};
        \node (a3) at (-5,0) {$...$};
        \node (a3) at (-5,0) {$...$};
        \node (a4) at (5,0) {$...$};
        \node (b0) at (0,-0.8)[font=\tiny] {$(k+1,k+1)$};
        \node (b1) at (2.5,-0.8) [font=\tiny]{$(k,k+1)$};
        \node (b2) at (-2.5,-0.8) [font=\tiny]{$(k+1,k)$};
        \node (b3) at (-5,-0.8) {$...$};
        \node (b4) at (5,-0.8) {$...$};
        \node at (0,1){$\vdots$};
        \node at (0,-1.5){$\vdots$};

    \end{tikzpicture}
    \caption{Structure of dynamic programming matrix with size $2\times (2q+1)$. Each element of the matrix is a bitarray of vertex in graph.}
    \label{fig:dpMatrix}
\end{figure}
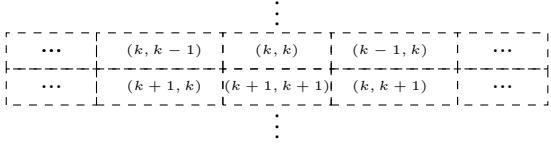
 
So we have Algorithm\ref{alg:ECG} in appendix. 
 
The following Figure\ref{fig:ECGflowChart} is a flowchart illustrating the iteration process of ECG at a specific step.
\begin{figure}[htbp]  
    \centering 
      \includegraphics[scale=0.5]{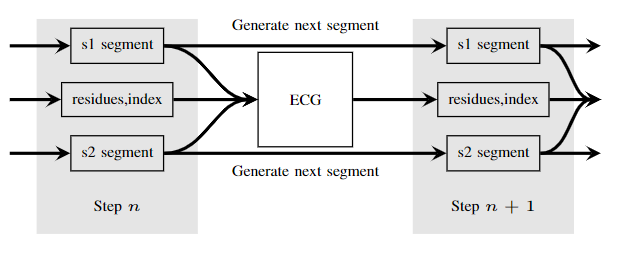}
    \caption{The flow chart of ECG iteration process.}
    \label{fig:ECGflowChart}
\end{figure}

\section{Scheme of Codebook Generation}
\label{sec:CodebookGeneration}

The performance of this scheme relies heavily on optimization and candidate generation. To demonstrate that the scheme can accomplish the codebook generation task, only Monte Carlo methods were employed for implementation: generate several suffixes at random and select the option with the least loss.

Based on \cite{Liu_He_Tang_2022}, we present Algorithm\ref{al:CG} here as a reference, ensuring that each iteration's candidate maintains a certain GC-balance and run length.

\section{Analysis}
\label{sec:analysis}
\subsection{Redundancy}
For whether error corrections can be achieved, we can calculate by constrained edit distance. Therefore, we mainly analyze redundancy.

We have 
\[n(C) = \log_2(\lvert C\rvert ).\]
And the redundancy of $C$ is 
\[r(C) = n\log_2(q)-n(C).\]
Since our approach generates sequences based on the size of codebook $C$, when comparing with other methods, we adopt the codebook-redundancy comparison approach.

\subsection{Complexity}
The overall complexity is composed of two aspects
\begin{itemize}
    \item The computation of ECG.
    \item The computation of codebook generation.
\end{itemize}
Fixed the number of $m=|C|$, the required codebook $C$ grows iteratively step by step. In this process, ECG needs to be computed independently $\frac{m(m-1)}{2}$ times. The steps of candidate generation and optimal selection heavily depend on the specific implementation, and therefore, this part is not within the scope of the discussion.

Next, we will focus our attention on the analysis of ECG.

For space complexity, we use a $2\times (2q+1)$ matrix to preserve the intermediate result, where $q$ is the maximal number of ins/del. The size of each element of the matrix depends on the setting of $EQ$, which is $L/8$, $L=\prod_{i}(EQ[i]+1)$. Thus, the space complexity is at least $O(qLm^2)$.

In terms of time complexity, let's first count how many nodes need to be traversed,
\[\sum_{i=0}^q (2i+1) + (2q+1)(n-q-1)=n(2q+1)-q(q+1).\]
The number of transitions (Algorithm\ref{alg:TransOfFEC}) each node needs to perform is $|EQ|$, and the complexity of single transition is $O(3\lceil L/|Byte|\rceil)=O(L)$ (Bit operations are parallel for each bit in a byte). Hence, the computational complexity for a single ECG calculation is at most 
\[O((n(2q+1)-q(q+1))(|EQ|L))=O(nqL|EQ|),q\ll n,\]
and the minimum complexity for one iteration of codebook generation is $O(m^2n(2q+1)-q(q+1))$.

\subsection{Result}
Experimental data were obtained with the following equipment Table\ref{tab:dev}.

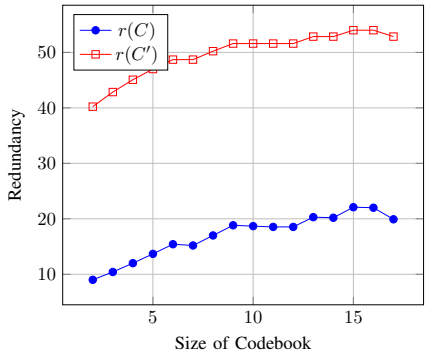
\begin{figure}[htbp]
    \centering
    \begin{tikzpicture}[scale=0.7]
  \begin{axis}[
    xlabel={Size of Codebook},
    ylabel={Redundancy},
    grid=major,
    legend pos=north west,
  ]

  \addplot[mark=*,blue] coordinates {
    (2, 9)
    (3, 10.4150375)
    (4, 12)
    (5, 13.67807191)
    (6, 15.4150375)
    (7, 15.19264508)
    (8, 17)
    (9, 18.830075)
    (10, 18.67807191)
    (11, 18.54056838)
    (12, 18.54056838)
    (13, 20.29956028)
    (14, 20.19264508)
    (15, 22.0931094)
    (16, 22)
    (17, 19.91253716)
  };
  \addlegendentry{$r(C)$}

  \addplot[mark=square,red] coordinates {
     (2, 40.21928095)
    (3,  42.84962501)
    (4,  45.07354922)
    (5, 47)
    (6, 48.69925001)
    (7, 48.69925001)
    (8, 50.21928095)
    (9, 51.59431619)
    (10, 51.59431619)
    (11, 51.59431619)
    (12, 51.59431619)
    (13, 52.84962501)
    (14, 52.84962501)
    (15, 54.00439718)
    (16, 54.00439718)
    (17, 52.84962501)
  }; 
  \addlegendentry{$r(C')$}

  \end{axis}
\end{tikzpicture}
    \caption{Comparison graph with state-of-the-art results \cite{Smagloy_Welter_Wachter-Zeh_Yaakobi_2020}. Our approach computes three times for different codebook sizes and returns the length of the sequence $n$. The state-of-the-art result is $10log(n) + 3log(q) + 11,q=4$.}
    \label{fig:Comp1}
\end{figure}

We conducted tests on the codebook for correcting single-substitution  single-deletion errors, comparing the results with the state-of-the-art outcomes\cite{Smagloy_Welter_Wachter-Zeh_Yaakobi_2020}. The comparison results are illustrated in the Figure\ref{fig:Comp1}.

Because the actual runtime varies depending on the specific candidate selection and optimization strategies, here, only a reference is provided Table\ref{tab:timeConsumpStep}.
 
For additional results, please refer to \href{https://github.com/ylu1997/Code_Generation_For_DNA_Storage}{our GitHub}.
\section{Conclusions}
\label{sec:Conclu}

In conclusion, this study has tackled the challenges associated with error correction codes for DNA storage by transforming them into a problem analogous to spell correction in natural language processing (NLP). By presenting a novel approach to construct a codebook, the complexities of efficient encoding, decoding, and error correction are alleviated. What was once a formidable task of generating a highly redundant codebook is now reframed as a puzzle akin to playing Go, offering a plethora of methods for resolution, reminiscent of the strategies employed by AlphaGo. In the scenario of error correction requirements with low hardware dependence, such as error correction codes for DNA storage, a general approach is proposed, eliminating the need to spend time constructing algorithms.

The proposed algorithm successfully mitigates the obstacles associated with individual bit consideration. However, it is important to note that challenges persist when addressing burst errors, and further discussion and exploration are warranted in this context. Future research endeavors should delve into refining the algorithm to encompass and effectively handle scenarios involving burst errors. This advancement will contribute to the continued evolution of DNA storage technologies, ensuring their robustness and effectiveness in diverse contexts.

\newpage
\bibliography{referencs}
\bibliographystyle{IEEEtran}

\newpage
\appendices

\section{ }
\label{sec:appendix} 
\begin{algorithm} 
\caption{Transition of FEC}
\label{alg:TransOfFEC} 
\KwData{FEC bit array of the source node$\beta$; Type of edge $T$; FEC bit array of the target node $\beta'$} 
\KwResult{FEC bit array of the target node $\beta'$} 
 Obtain the indices $I\subseteq [|EOI|]$ of all edits belonging to the class $T$\;
 \If{$T!=\epsilon$}{
     Initiate a bit array $\alpha$\;
     \For{$i \in I$}{
        $\alpha\leftarrow bitAnd(\beta,EM[i])$\;
        $\alpha\leftarrow ShiftHight(\alpha, SA[i])$\;
        $\beta'\leftarrow bitOr(\beta',\alpha)$\;
     }
 }
 \Else{
    $\beta'\leftarrow \beta$\;
 } 
\end{algorithm} 

\begin{algorithm}
\caption{Algorithm ECG}
\label{alg:ECG}
\KwData{$EOI,EQ,EM,SA,q$}
\KwIn{String $s_1,s_2\in \Alphabet{4}^n$;Start index of string $i_0$; End index of string $i_1$; Last residue $m$; Start index of row $r_0$}
\KwResult{Residue $m'$; Next index of row $r_1$}
$m'=Deepcopy(m)$ \;
\For{$i\in \{i_0,...,i_1-1\}$}{
    $d\leftarrow min\{q,i\}$\;
    \For{$j\in\{-2d,...,1\}$}{
        $idx1 \leftarrow (i+\lfloor j/2 \rfloor) * (j \mod{2})$\;
        $idx2 \leftarrow (i+\lfloor j/2 \rfloor) * ((j+1) \mod{2})$\;
        $bias \leftarrow \lfloor j/2 \rfloor * (-1)^{(j \mod{2})}$;$row \leftarrow i-i_0+r_0$;$col \leftarrow q+bias$\;
        \If{$s_1[idx1]==s_2[idx2]$}{
            \tcp{Trans is Algorithm\ref{alg:TransOfFEC} (source node, edge type, target node)}
            $Trans(m[row-1,col],\epsilon,m[row,col])$ \;
        }
        \Else{
            $tmp \leftarrow AllZeroBitArray\in \Alphabet{2}^L$\;
            \If{  $ (s_{1} [idx1] \ne \text{' '})\&\& (s_{2} [idx2] \ne \text{' '})$}{
                \tcp{Substitution}
                $Trans(m[row-1,col],\subOp,tmp)$ \;
            }
            \If{$(idx1 != 0)\&\&(col < 2q)$}{
                \tcp{Deletion}
                $Trans(m[row + (-1 \text{ if } bias < 0 \text{ else } 0), col + 1],\delOp,tmp)$ \;
            }
            \If{$(idx2 !=0)\&\&(col > 0)$}{
                \tcp{Insertion}
                $Trans(m[row + (-1 \text{ if } bias >0 \text{ else } 0), col - 1],\insOp,tmp)$ \;}
        }
        $m'[row,col]\leftarrow tmp$\;    
    }
}
$r_{1}\leftarrow r_{0} + i_{1} - i_{0}$ \;
\end{algorithm}  
\begin{algorithm}
  \caption{Candidate Generation}
\label{al:CG}
  \KwData{$s,run,bal,L_1,L_2$}
  \KwResult{$C$}
  $capGC =  bal * (L_1+L_2)$\;
  Initialize a list $A=\{s[n-L_1:]\}$ and empty list $B=[]$\;
   
  \While{$A\ne \emptyset$}{
    $item = A.pop(0)$\;
	\If{$item.len()==L_2$}{
		$C.append(item)$\;
		continue\;
	}
    $itemLastNuc = item[-1]$\;
    $itemLastRun = item.lastRun()$ the run number of the last nucleotide\;
    $itemGCcount = item.countGC()$ \;
    \For{$tail \in S$}{
        $tailRun=tail.len()$\;
        $tailNuc=tail[0]$\;
		 \If{$tailNuc == G $ or $  C$ and $itemGCcount+tailRun>capGC$}
			{continue\;}
		 \If{$tailNuc == A $ or $  T$ and $L_1-itemGCcount+tailRun>L_1+L_2-capGC$}
			{continue\;}
		 \If{$lastNuc==itemLastNuc$ and $itemLastRun+tailRun\ge run$}{
           continue\;}
		$B.append(concatenate(item,tail))$\;
    }
    $A=B$\;
    $B=\emptyset$\;
  }

\end{algorithm}  

\begin{table}[htbp]
    \centering
    \begin{tabular}{c|ccccc}
        \hline
         Augmental Length& 2& 3&4&5&8 \\
         \hline
         Time Consumption(s)& 35.72&29.23&29.39&25.21&24.82\\
         \hline
         Sequence Length & 16 & 15& 16& 15&16\\
         \hline
         Redundancy & 27&25 &27 & 25&27\\
         \hline
    \end{tabular}
    \caption{The time consumption table regarding the generation of a codebook capable of handling single-substitution single-deletion.}
    \label{tab:timeConsumpStep}
\end{table}

\begin{table}[t]
  \centering
    \begin{tabular}{c|p{5cm}<{\centering}}
        Processor:& Intel(R) Core(TM) i7-7700HQ CPU @ 2.80GHz 2.80 GHz.\\
        Memory (RAM):& 16.0 GB (15.9 GB available). \\
        System Type:& 64-bit operating system, x64-based processor.\\
        Python Version:& 3.7.3\\
        Number of threads used&1.
    \end{tabular}
    \caption{Device Parameters}
    \label{tab:dev}
\end{table}

\end{document}